\documentstyle[amsfonts,aps,twocolumn]{revtex}

  2
 1

\newcommand{\eh}{\mathcal{E}(\mathcal{H})}

\begin{document}
\title{Quantum states and generalized observables:\\
a simple proof of Gleason's theorem}
\author{P.~Busch\thanks{Electronic address: p.busch@hull.ac.uk}}
\address{{\small Department of Mathematics, University of Hull, Hull HU6 7RX, UK}}

\date{{\small 28 May 2003}}
\maketitle

\begin{abstract}
\noindent A quantum state can be understood in a loose sense as a map that
assigns a value to every observable. Formalizing this characterization of
states in terms of generalized probability distributions on the set of
\emph{effects}, we obtain a simple proof of the result, analogous to Gleason's
theorem, that any quantum state is given by a density operator. As a corollary
we obtain a von Neumann-type argument against non-contextual hidden variables.
It follows that on an {\em individual} interpretation of quantum mechanics, the
values of effects are appropriately understood as {\em propensities}.

\vspace{12pt}

\end{abstract}

\vspace{12pt}

In this paper we will characterize a notion of quantum states that takes into
account the general representation of observables as `positive operator valued
measurements' (POVMs). The idea of a state as an expectation value assignment
will be extended to that of a generalized probability measure on the set
$\mathcal{E}(\mathcal{H})$ of all \emph{effects}, that is, the positive
operators which can occur in the range of a POVM \cite{BGL}. All such
generalized probability measures are found to be of the standard form, i.e.,
determined by a density operator. This result constitutes a simplified proof
and at the same time more comprehensive variant of Gleason's theorem
\cite{Glea}. The paper concludes with an application of this result to the
question of hidden variables \cite{Busch99}.

In the traditional formulation of quantum mechanics in Hilbert space, states
are described as density operators and observables are represented as
self-adjoint operators. Alternatively, and equivalently, experimental events
and propositions are represented as orthogonal projection operators, and states
are  defined as generalized probability measures on the non-Boolean lattice
$\mathcal{P}(\mathcal{H})$ of projections, i.e., as functions
$E\mapsto v(E)$ with the properties\\
(P1) $0\le v(E)\le 1$ for all $E$;\\
(P2) $v(I)=1$;\\
(P3) $v(E+F+\dots)=v(E)+v(F)+\dots$ for any sequence $E,F,\dots$
with $E+F+\dots\le I$.\\
According to Gleason's theorem \cite{Glea}, all states are given by density
operators so that $v(E)=v_\rho(E)=tr[\rho E]$, provided that the dimension of
the complex Hilbert space is at least 3. The duality of states and observables
is thus characterized through the trace expression $tr[\rho E]$, which in the
minimal interpretation gives the probability of an outcome associated with $E$,
of a measurement performed on a system in state $\rho$.

In quantum physics there are many experimental procedures leading to
measurements whose outcome probabilities are expectations not of projections
but rather of \emph{effects}. It is therefore natural to define a quantum state
as a generalized probability measure not just on $\mathcal{P}(\mathcal{H})$ but
on the full set of effects, $\mathcal{E}(\mathcal{H})$, in such a way that the
conditions $(P1)-(P3)$ hold for all $E,F,\dots\in\mathcal{E}(\mathcal{H})$.
Note that while for sets of projections the condition
$E+F+\dots\in\mathcal{P}(\mathcal{H})$ is equivalent to $E,F,\dots$ being
mutually orthogonal and thus commuting, the commutativity is no longer
necessary for $E+F+\dots\le I$ to hold if $E,F,\dots$ are effects. The
following analogue of Gleason's theorem then holds.

\noindent{\bf Theorem.} Any generalized probability measure $E\mapsto v(E)$ on
$\mathcal{E}(\mathcal{H})$ with the properties $(P1)-(P3)$ is of the form
$v(E)=tr[\rho E]$ for all $E$, for some density operator $\rho$.

\noindent {\em Proof.} It is trivial to see that $v(E)=nv(\frac{1}{n}E)$ for
all positive integers. Then it follows immediately that $v(pE)=pv(E)$ for any
rational $p\in \lbrack 0,1]$. Observe also the additivity and positivity entail
that any effect valuation is order preserving, $E\leq F\Longrightarrow v(E)\leq
v(F)$. Let $\alpha $ be any real number, $0\leq \alpha \leq 1$. Let $p_{\mu }$
and $q_{\nu }$ be sequences of rational numbers in [0,1] such that $p_{\mu
}\nearrow \alpha $ and $q_{\nu }\searrow \alpha $. It follows that $v(p_{\mu
}E)=p_{\mu }v(E)\leq v(\alpha E)\leq v(q_{\nu }E)=q_{\nu }v(E)$. Hence,
$v(\alpha E) =\alpha v(E)$.

Let $A$ be any positive bounded operator not in ${\cal E({H})}$. We can always
write $A=\alpha E$, with $E\in {\cal E({H})}$ and suitable $\alpha  \ge 1$. Let
$E_1,E_2\in {\cal E({H})}$ be such that $A=\alpha _1E_1=\alpha _2E_2$. Assume
without loss of generality that $1\le \alpha_1<\alpha _2$. Then
$v(E_2)=\frac{\alpha _1}{\alpha _2}v(E_1)$, and so $\alpha _1 v(E_1) = \alpha
_2v(E_2)$. Thus we can uniquely define $v(A)=\alpha _1v(E_1)$.

Let $A,B$ be positive bounded operators. Take $\gamma >1$ such that
$\frac{1}{\gamma }(A+B) \in {\cal E({H})}$. Then we can write $v(A+B)$ as
$\gamma v({{\textstyle{\frac{1}{\gamma }}}}(A+B))= \gamma
v({{\textstyle{\frac{1}{\gamma }}}}A)+\gamma v({{\textstyle{\frac{1}{\gamma
}}}}B)=v(A)+v(B)$.

Finally, let $C$ be an arbitrary bounded self-adjoint operator. Assume we have
two different decompositions $C=A-B=A^{\prime }-B^{\prime }$ into a difference
of positive operators. We have $v(A)+v(B^{\prime })=v(B)+v(A^{\prime })$ and so
$v(A)-v(B)=v(A^{\prime })-v(B^{\prime })$. Thus we can uniquely define:
$v(C):=v(A)-v(B)$. It is now straightforward to verify the linearity of the map
$v$ thus extended to all of ${\cal L} _{s} \left( {\cal H}\right) $. We have
found that any generalized probability measure on effects extends to a unique
positive linear functional which is normal (due to the $\sigma $-additivity).
It is well known that any such functional is obtained from a density operator
(e.g., \cite{Dav}, Lemma 1.6.1, or see the direct elementary proof due to von
Neumann \cite{vN,BelBug}).\ $\square$

The conclusion of our theorem is the same as that of Gleason's theorem. The
extreme simplicity of the proof in comparison to Gleason's proof is due to the
fact that the domain of generalized probability measures is substantially
enlarged, from the set of projections to that of all effects.

The statement of the present theorem also extends to the case of 2-dimensional
Hilbert spaces where Gleason's theorem fails. It is worth noting that the
dispersion-free valuations constructed on the set of projections of a
2-dimensional Hilbert space (see, e.g., \cite{Bell,KS}), simply do not extend
to any valuations on the full set of effects. The reason must be seen in the
fact that the additivity requirement for $v$ on sets of pairwise orthogonal
projections is too weak to enforce the linearity of $v$, considering that such
sets of projections are mutually commutative.

Here is a simple intuitive argument demonstrating that there are no linear
extensions of any dispersion-free valuation on the projections of a
2-dimensional Hilbert space. We use the Poincar\'e sphere representation of
positive operators of trace 1, $A=1/2\,(I+\mathbf a\cdot\sigma)$, where
$\sigma=(\sigma_x,\sigma_y,\sigma_z)$, ${\mathbf a}=(a_x,a_y,a_z)$, with
$|{\mathbf a}|^2=a_x^2+a_y^2+a_z^2\le 1$. All projections are then either $I$
or $O$ or $P=1/2\,(I+\mathbf n\cdot\sigma)$, with $|\mathbf n|=1$. Let $v$ be a
dispersion-free valuation on the projections. Any pair of mutually orthogonal
projections $P,P'=I-P$ will have values 1 and 0 such that their sum is 1. Hence
there are non-orthogonal pairs $P=1/2\,(I+\mathbf n\cdot\sigma)$, $Q=
1/2\,(I+\mathbf m\cdot\sigma)$ such that both have value 0. If $v$ had a linear
extension, then all the effects corresponding to the line segment joining
$\mathbf n$ and $\mathbf m$, $E=\lambda P+(1-\lambda)Q$, with $0\le\lambda\le
1$, would have values $v(E)=\lambda v(P)+(1-\lambda)v(Q)=0$. On the other hand,
we can write $E$ in its spectral decomposition $E=\mu R+(1-\mu)R'$, where
$0<\mu<1$ if $0<\lambda<1$. Assume that $v(R)=1$ $v(R')=0$, then $v(E)=\mu\ne
0$, which contradicts the previous conclusion that $v(E)=0$. Hence there is no
consistent linear extension of $v$.

Up to this point we have restricted ourselves to the minimal interpretation of
quantum states and observables, according to which these entities are tools for
calculating experimental probabilities. We have shown that, given the set of
effects as a representation of all experimental yes-no questions, any quantum
state, understood as a generalized probability measure on the set of effects,
is given in the familiar way by a density operator.

This result entails a formalization of the well-known fact that quantum
mechanics is an irreducibly probabilistic theory: in contrast to classical
probability theory, quantum probabilities cannot be decomposed into convex
combinations of dispersion-free (that is, $\{0,1\}$-valued) generalized
probability measures.

We conclude with a brief outline of an application of the above result to
interpretations of quantum mechanics that go beyond the scope of the minimal
interpretation. Such interpretation will consider observables as
representations of \emph{properties} of a system and effects as yes-no
propositions about the possible values of the observables. The role of states
will be to assign values to observables and effects. In a deterministic world,
one would expect a complete state description to assign one of the values 1 or
0 to each effect of a complete collection $E_i$ (with $\sum E_i=I$), in such a
way that 1 occurs exactly once. Thus the sum of the values for all $E_i$ is 1.

This consideration leads to the idea of defining states as \emph{effect
valuations}, that is, as functions $v:E\mapsto v(E)$ of effects with the
properties:  $v(E)\ge 0$, and
 $v(E)+v(F)+\dots=1$ if $E+F+\dots=I$.

It is easy to see that every effect valuation has the properties (P1-3) of
generalized probability measures, and conversely. Hence the above theorem
entails that any effect valuation is of the form $v(E)=tr[\rho E]$ for all
$E\in\mathcal{E}(\mathcal{H})$ and some density operator $\rho$.

An interpretation of valuations as truth value assignments would require the
numbers $v(E)$ to be either 1 or 0, indicating the occurrence or nonoccurrence
of an outcome associated with  $E$. Valuations with this property are referred
to as \emph{dispersion-free}. The above theorem entails immediately that
dispersion-free effect valuations which are defined everywhere on $\eh$ do not
exist. It follows that non-contextual hidden variables, understood as
dispersion-free, globally defined, valuations, are excluded in quantum
mechanics.

The argument against non-contextual hidden variables thus obtained resembles
formally that of von Neumann \cite{vN}. However, von Neumann's problematical
assumption, that of additivity of a valuation over arbitrary (countable) sets
of (commuting or noncommuting) self-adjoint operators \cite{JamMer}, is here
replaced by the requirement of additivity over (countable) sets of effects that
add up to $I$. Such collections of effects constitute a POVM and are thus
jointly measurable in a single experiment. It makes thus sense to consider
hypothetical simultaneous (hidden, dispersion-free) values of such sets of
effects, and hence also the values of sums of effects provided these sums are
bounded by $I$.

In the case of a pure state $\rho=|\varphi\rangle\langle\varphi|$, the
occurrence of values $v_\rho(E)$ strictly between 0 and 1 indicates a situation
where the property associated with $E$ is objectively indeterminate, that is
its presence or absence is not just subjectively unknown. This interpretation
is in accord with the \emph{propensity} interpretation of probabilities,
according to which the number $v_\rho(E)$ gives a measure of the system's
objective tendency to trigger an outcome represented by effect $E$ if the state
is given by $\rho$ and a measurement is made of a POVM containing $E$
\cite{Gisin}.

As an example, $E_1$ and $E_2=I-E_1$ could represent the propositions that a
quantum particle is in the upper and lower path of an interferometer,
respectively. If a pure state $\rho$ is a superposition of states $\rho_1$,
$\rho_2$ in which $E_1$ and $E_2$ are real, respectively (i.e.,
$v_{\rho_1}(E_1)=v_{\rho_2}(E_2)=1$), then there is no convex decomposition of
that state in terms of valuations which are dispersion-free, even only with
respect to $E_1,E_2$. The fact that $0<v_\rho(E_i)<1$ is then an expression of
the \emph{indeterminateness} of the properties $E_1,E_2$ in the state $\rho$.
The most appropriate way of accounting for this situation seems to be to say
that the localization of the quantum particle is extended over the space
occupied by the two paths of the interferometer. The quantum particle is
present, \emph{to a degree quantified by the number} $v_\rho(E_i)$, in each of
the two paths represented by $E_i$. If forced by a measurement to decide
whether to show up in the upper or lower path, it will do so with a propensity
quantified by those numbers.

A related interpretation of valuations for unsharp measurements as approximate
truth values has recently been advocated by T.~Breuer in this journal
\cite{Breu}, who applied Gleason's theorem to obtain a Kochen-Specker theorem
\cite{KS} for unsharp spin observables.

The nonexistence of dispersion free effect valuations raises the interesting
question whether there are subsets in the set of effects, with meaningful
structures, on which such dispersion-free valuations can be defined.
Interesting constructions demonstrating a positive answer to this question are
presented for subsets of projections in \cite{Bub}, or also for effects in
\cite{Kent}. Intuitively, it appears that the valuations of Bub \cite{Bub} are
defined on relatively sparse sets of projections, but these sets do possess
some structures that can be argued to be necessary for a consistent set of
definite properties; by contrast, the valuations of Kent \cite{Kent} are
defined on `dense' sets of POVMs where it is not obvious that these are
equipped with such `logical' structures. The important task remains to explore
how far one can go in defining non-contextual dispersion-free valuations on
subsets of effects with appropriate structures, without running into conflict
with the modified Gleason theorem proven here.

\end{document}